# State and Parameter Estimation of
# The Lorenz System
# In Existence of Colored Noise

**Mozhgan Mombeini**[a]
**Hamid Khaloozadeh**[b]

[a] Electrical Control and System Engineering, Researcher of Institute for Research in Fundamental Sciences (IPM), Tehran, Iran

[b] Faculty of Electrical & Computer Eng., K.N. Toosi Univ. of Technol., Tehran, Iran

E-mail: mzh.mombeini@ipm.ir[a] , H_Khaloozadeh@kntu.ac.ir[b]

**Abstract:** Many researchers are interested to use Extended Kalman Filter (EKF) for state estimation of complex nonlinear dynamics with uncertainties which modeled with white noises. On the other hand behavior of the chaotic systems in time domain itself is similar to noise too. In this paper, states of the chaotic Lorenz system that its uncertainties modeled with colored noise on states and also on output are considered. For both cases, the case that parameters of Lorenz system are known and the case that parameters of the Lorenz system are unknown, EKF is used to estimate the states. In the case that parameters are unknown using a stochastic viewpoint parameters of the system and parameters of the first order filter of the colored noise are estimated. Efficiency of the method is shown with simulation.

## 1. Introduction

Extended Kalman Filter (EKF) is a powerful tool for estimation of nonlinear systems states in presence of noise and used in many nonlinear cases. In some nonlinear systems by changing parameters of the systems they get to period doubling and continuing the parameter changing then system becomes chaotic. This procedure determines the domain in which system is chaotic, where behavior of the system in time domain is similar to noise and also system is sensitive to initial condition. Then according the noise like behavior of the chaotic systems, efficiency of the EKF for state and parameter estimation of these systems is investigated in some works [1-3].
In this paper, efficiency of the EKF for separation of chaotic states from noise is investigated for Lorenz system with known parameters and for Lorenz system with unknown parameter. Where, in both cases the existent noise is a colored noise and EKF is used to estimate the filter parameters too.

## 2. Discrete modeling and different uncertainties

For continuous model of system in state space with

$$\dot{x} = f(x, \theta)$$
$$z = Hx(t)$$

Where $x \varepsilon R^n$ denoted to state variables $\theta$ denote to system parameters, $z \varepsilon R^m$ is output and $H$, measurement matrix is a $m \times n$ matrix, a forward difference approximation for derivations of the states is $\dot{x} = \frac{x(t + \Delta t, \theta) - x(t, \theta)}{\Delta t}$. Choosing $\Delta t$ small enough as sample time, $x_{k+1} = x(t + \Delta t, \theta)$ and $x_k = x(t, \theta)$ result in $\frac{x_{k+1} - x_k}{\Delta t} = f(x_k, \theta)$ or equivalently a discrete model as

$$x_{k+1} = F(x_k, \theta)$$

which can be simulated with computer codes. Note that for $x \varepsilon R^n$ the forward approximation of derivatives of each state should be calculated separately. Then realization of discrete model with measured output $z \varepsilon R^m$ in state space is

$$x_{k+1} = F(x_k, \theta)$$
$$z = Hx_k \qquad (1)$$

### 2.1 White noise type uncertainty

White noises are some uncertainties that are independent time series and distributed identically, which means no auto correlation between them. In particular case, the "Gaussian white noise" has normal distribution with zero mean and standard variation $\sigma$. For system(1) with Gaussian white noise $\{w\}$ on states and Gaussian white noise $\{v\}$ on output it change as following

$$x_{k+1} = F(x_k, \theta) + Gw$$
$$z = Hx_k + v . \qquad (2)$$

Where, mean vector and autocorrelation matrix of $w$ and $v$ are $\mu_w = E\{w\} = 0$, $R_{ww} = E\{ww^T\} = \sigma_w^2 I$, $\mu_v = E\{v\} = 0$ $R_{vv} = E\{vv^T\} = \sigma_v^2 I$, $E$ is expected value operator, $I$ is identity matrix and $G$ is a matrix or vector that its dimension matches to size of other elements of the equation.

## 2.2 Colored noise type uncertainty

Colored noises are some uncertainties that are dependent to their past states then have auto correlation. In particular case passing a "Gaussian white noise" from a first order filter results in a colored noise.

Model of a first order filter which result in $\{w\}$ colored noise on states is

$$w_{k+1} = cw_k + e^1 \qquad (3)$$

Where, $\{e^1\}$ is Gaussian white noise and $c$ is a real constant. Similarly by defining

$$v_{k+1} = dv_k + e^2 \qquad (4)$$

Where $\{e^2\}$ is Gaussian white noise and $d$ is a real constant, $\{v\}$ is realization for colored noise on measured output.

## 2.3 Uncertain parameters

For most mathematical modeling of systems there are uncertainties in parameters of the model. Some are inherent result of the accuracy in modeling some are because of variation in realistic condition of the system among the time and etc. Then, although the $\theta$ parameters of the system (1), are considered constant in observation scale time but they may have changes among the time. In other word in bigger scale times these uncertain parameters are variable. Here, using stochastic perspective, for a unknown parameter $\bar{\theta}$ by variation in the range $[\theta - \Delta\theta, \theta + \Delta\theta]$, the parameter is consider the as a more general case a white noise with $\theta$ mean value and $\Delta\theta$ variance value or equivalently

$$\begin{aligned} \bar{\theta} &= \theta + \vartheta \\ \vartheta &\sim N(0, \Delta\theta) \end{aligned} \qquad (5)$$

## 3. Extended Kalman Filter (EKF)

Extended Kalman Filter is a mathematical tool for state estimation of nonlinear systems and the method is used in many cases [][]. It called First-Order Filter too, because it uses an approximation of the system with expanding it in Taylor series getting the first order terms, and higher order terms (H.O.T.) are considered negligible. In short description, for the nonlinear model with Gaussian noise

$$\begin{aligned} x_{k+1} &= F(x_k, \theta) + G(x_k, w) \\ y &= H(x_k) + v \, . \\ w &\sim N(0, Q), \, v \sim N(0, R), \, x_0 \sim N(\bar{x}_0, Q_0) \end{aligned} \qquad (6)$$

Where $x \varepsilon R^n$ denoted to state variables $z \varepsilon R^m$ is output and $v$, $w$ are white noises with zero mean and standard deviation $Q$, $R$, linearization of (6) at every step $k$ results in $\overline{A}_k = \left.\dfrac{\partial F(x_k,\theta)}{\partial x_k}\right|_{x_k}$, $\overline{C}_k = \left.\dfrac{\partial H(x_k)}{\partial x_k}\right|_{x_k}$, $\overline{G}_k = G(x_k)$,

Extended kalman filter recursively estimates the states in two phases, forecast and correction.

### 3.1 EKF for state estimation

For system (6) with $x_k$ states and $\theta$ known constant parameters forecast phase is

$$\hat{x}^-_{k+1} = F(\hat{x}_k, \theta)$$
$$P^-_{k+1} = \overline{A}_k P_k \overline{A}'_k + \overline{G}_k Q \overline{G}'_k \qquad (7)$$
$$\hat{x}_0^- = \overline{x}_0, P_0^- = Q_0$$

And correction phase is

$$\hat{x}_k = \hat{x}_k^- + L_k(y_k - H(\hat{x}_k^-))$$
$$L_k = P_k^- \overline{C}'_k (\overline{C}_k P_k^- \overline{C}'_k + R)^{-1} \qquad (8)$$
$$P_k = P_k^- - L_k \overline{C}_k P_k^-$$

Where $\hat{x}_k$ is the estimated value for the states.

### 3.2 EKF for state and parameters estimation

A method is introduced in literature to extend application of the EKF method for estimation of uncertain parameters too. Essence of the method is to consider the unknown parameter $\theta$ as an additional state [1-3].

In this paper, the method is used for parameter estimation of the discrete model. Consider the discrete model (2) with $\theta$ uncertain parameters. By considering this uncertainty on parameters as a noise model (5) and defining augmented vector for new states as $\zeta = [x, \theta]$, state space model (2) changes to following equation

$$\xi_{k+1} = \begin{bmatrix} x_{k+1} \\ \theta_{k+1} \end{bmatrix} = F'(\xi) + G'W' = \begin{bmatrix} F(x_k, \theta_k) \\ \theta_k \end{bmatrix} + \begin{bmatrix} G & 0 \\ 0 & I \end{bmatrix} \begin{bmatrix} w \\ \vartheta \end{bmatrix} \quad (9)$$

$$y = Hx + v$$

$$w \sim N(0,Q), v \sim N(0,R), \vartheta \sim N(\bar{\theta}, \Delta\theta)$$

This new description of the system with unknown parameters is in the class of system (6) and EKF method can be used for new states $\zeta = [x, \theta]$ with linearization of (9) at every sampling time instant $k$.

## 4. Effect of uncertainty on Lorenz system

The Lorenz system is a chaotic system that introduced in 1963 with equations

$$\dot{x}_1 = \sigma(x_2 - x_1)$$
$$\dot{x}_2 = -x_1 x_3 + r x_1 - x_2 \quad (10)$$
$$\dot{x}_3 = x_1 x_2 - b x_3$$

The system is chaotic in range of parameters, here considered with $\sigma = 10, b = 1.25, r = 28$.

Discrete model (1) for system (10) is

$$x_{1k+1} = x_{1k} + \Delta t \sigma(x_{2k} - x_{1k})$$
$$x_{2k+1} = x_{2k} + \Delta t(-x_{1k} x_{3k} + r x_{1k} - x_{2k}) \quad (11)$$
$$x_{3k+1} = x_{3k} + \Delta t(x_{1k} x_{2k} - b x_{3k})$$

In this part, using simulation, effect of the different uncertainties on Lorenz system trajectory is shown. To quantify difference between states and estimation or perturbed states, error function which is used here is

$$error(x_n, y_n) = \frac{1}{N} \sum_{n=1}^{N} \|x_n - y_n\| \quad (12)$$

Where $x_n$ and $y_n$ are compared states, $N = \dfrac{T}{\Delta t}$ is total number of samples of the system evolution and $\|\cdot\|$ is norm-two operator.

To model effect of noise on states and parameters the values of G and H matrix of the model (2) are chosen as $G = \begin{bmatrix} 1 & 1 & 1 \end{bmatrix}^T$, $H = \begin{bmatrix} 1 & 0 & 1 \end{bmatrix}$ and $\Delta t = .01$, error function is calculated with $T = 7, dt = 0.01$ or equivalently $N = 700$. All models are run with same initial condition $(0.5622, 0.7893, 0.3509)$ for states.

### 4.1 Lorenz system with white noise uncertainty

Lorenz system (11) with white noise uncertainty changes to

$$x'_{1k+1} = x'_{1k} + \Delta t \sigma(x'_{2k} - x'_{1k}) + w$$
$$x'_{2k+1} = x'_{2k} + \Delta t(-x'_{1k}x_{3k} + rx'_{1k} - x'_{2k}) + w$$
$$x'_{3k+1} = x'_{3k} + \Delta t(x'_{1k}x'_{2k} - bx'_{3k}) + w \quad (13)$$
$$z = x'_{1k} + x'_{3k} + v$$
$$w \sim N(0,Q), v \sim N(0,R),$$

Figures (1), (2) show trajectories of the ideal model (11) and noisy model (13) with $Q = 0.001$ and $R = 0.001$ in time domain and phase domain respectively. Deviation in system trajectory with adding the white noise on states is obvious in these figures. According to figures adding white noise on the states, system's behavior changes from the ideal model which is without any noise. Value of the error (12) is $error(x, x') = 61.0554$ in simulation.

### 4-2 Lorenz system with colored noise

According to (3), (4) Lorenz system (11) with colored noise uncertainty changes to

$$x''_{1k+1} = x''_{1k} + \Delta t \sigma(x''_{2k} - x''_{1k}) + w_k$$
$$x''_{2k+1} = x''_{2k} + \Delta t(-x''_{1k}x_{3k} + rx''_{1k} - x''_{2k}) + w_k$$
$$x''_{3k+1} = x''_{3k} + \Delta t(x''_{1k}x''_{2k} - bx''_{3k}) + w_k$$
$$w_{k+1} = cw_k + e^1 \quad (14)$$
$$v_{k+1} = dv_k + e^2$$
$$z = x'_{1k} + x'_{3k} + v_k$$
$$e^1 \sim N(0,Q^1), e^2 \sim N(0,Q^2),$$

Where the parameters of first order filter are chosen as $c = 0.1$, $d = 0.2$, $Q^1 = 0.001$ and $Q^2 = 0.001$.

Figures (3), (4) show trajectories of the real model (11) and noisy model (14) in time domain and phase domain respectively. According to figures adding colored noise on the states, system's behavior changes from the ideal model (11) which is without any noise. Value of the error (12) is $error(x, x'') = 73.1981$ in simulation.

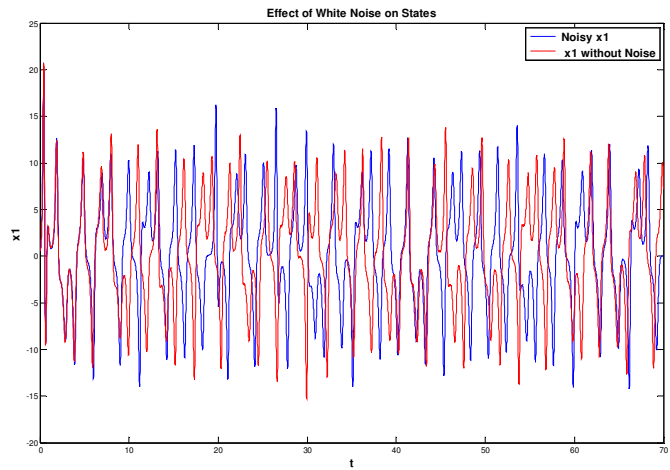

Fig.1 $x_1$ for ideal system (11) and $x_1'$ for noisy system (13)

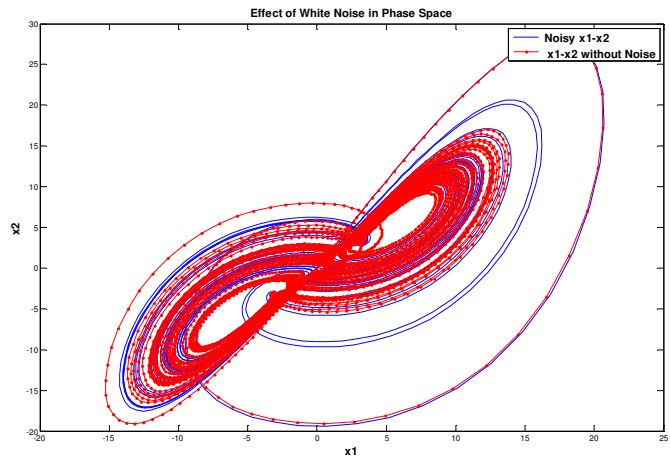

Fig.2 $x_1 - x_2$ for system (11) and $x_1' - x_2'$ for noisy system (13)

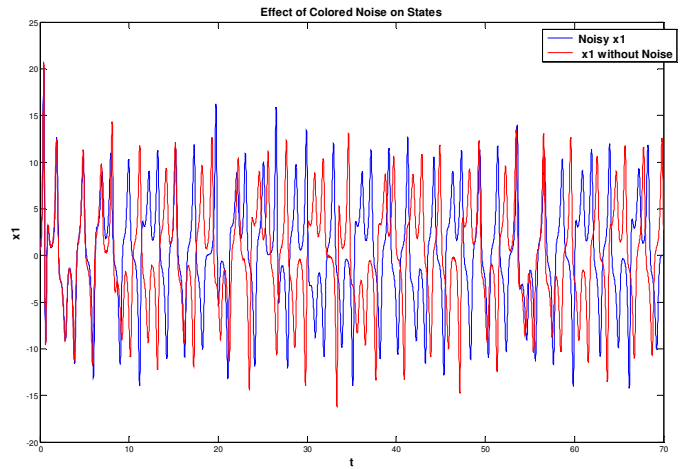

Fig.3 $x_1$ for ideal system (11) and $x_1''$ for noisy system (14)

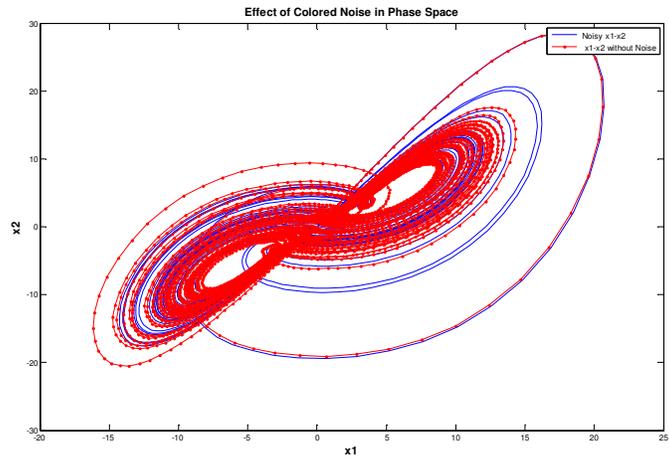

Fig.4 $x_1 - x_2$ for system (11) and $x_1'' - x_2''$ for noisy system (14)

### 4.3 Lorenz system with uncertain parameters

Although the chaotic systems are known with property of sensitivity to initial condition, this class of nonlinear systems is sensitive to parameters changes too. In this part this property is investigated on following uncertain Lorenz system

$$x'''_{1k+1} = x'''_{1k} + \Delta t \bar{\sigma}(x'''_{2k} - x'''_{1k})$$
$$x'''_{2k+1} = x'''_{2k} + \Delta t(-x'''_{1k} x'''_{3k} + \bar{r} x'''_{1k} - x'''_{2k}) \quad (15)$$
$$x'''_{3k+1} = x'''_{3k} + \Delta t(x'''_{1k} x'''_{2k} - \bar{b} x'''_{3k})$$
$$z = x'''_{1k} + x'''_{3k}$$
$$\bar{\sigma} = \sigma \pm \Delta\sigma, \bar{b} = b \pm \Delta b, \bar{r} = r \pm \Delta r, \bar{c} = c \pm \Delta c, \bar{d} = d \pm \Delta d$$

Where $\bar{\sigma}, \bar{b}, \bar{r}$ are nominal or mean value of the unknown parameters and $\Delta\sigma, \Delta b, \Delta r$ are range of the parameters variations.

Figures (5), (6) show trajectories of the model (15) with nominal parameters and with small variation of $\sigma$ parameter from $10$ to $10.001$ in time domain and phase domain respectively. According to figures with a small perturbation $\Delta\sigma = 0.005$, system's behavior changes from the nominal model. Value of the error (12) is $error(x, x''') = 65.6814$ in simulation.

### 5. EKF estimation of Lorenz system

In this part problem of estimation of different uncertain models are solved by changing them to model (6) which EKF is applicable for estimation of its states and parameters. For realization $x = [.]^T, \theta = [.]^T, Q' = [.]$ vectors are defined in this part. Where, $x$ is the state vector that EKF estimates, $\theta$ is vector of known parameters, $Q'$ is an auxiliary vector which its arrays are variance of the uncertainties that are considered as white noises in the system.

### 5.1 EKF for state estimation of Lorenz system with white noise on states

Realization of the Lorenz system (13) with white noise in the form (6) results in the following parameters

$$x = [x_1, x_2, x_3]^T, \theta = [\sigma, b, r]^T, Q' = [Q, R]$$

Figures (7), (8) show trajectories of the noisy model (13) and estimated value $\hat{x} = [\hat{x}_1, \hat{x}_2, \hat{x}_3]^T$ with EKF in time domain and phase domain respectively. According to figures EKF estimated the real value of $x$ correctly. Value of the error is $error(x, \hat{x}) = 0.0013$ in simulation.

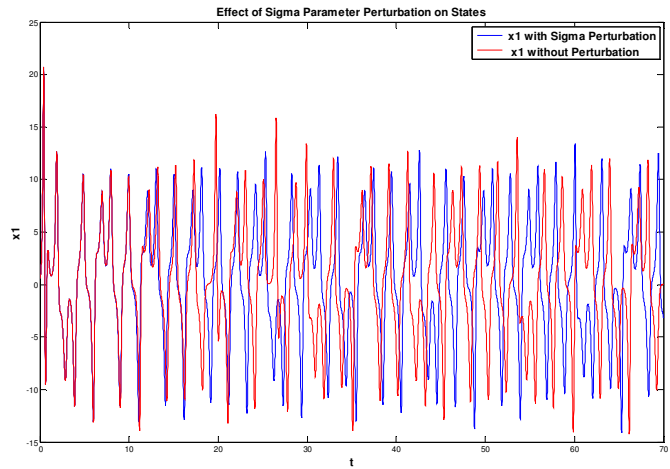

Fig.5 $x_1$ for ideal system (11) and $x_1'''$ perturbed system (15) with $\Delta\sigma = 0.001$

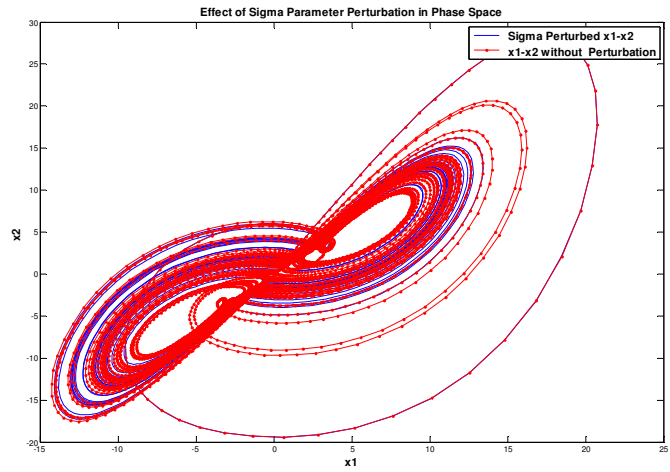

Fig.6 $x_1 - x_2$ for system (11) and $x_1''' - x_2'''$ for perturbed system (15) with $\Delta\sigma = 0.001$

## 5.2 EKF for state estimation of system with colored noise

Realization of the Lorenz system (14) with white noise in the form (6) results in the following parameters

$$x = [x'_1, x'_2, x'_3, v, w]^T, \theta = [\sigma, b, r, c, d]^T, Q' = [Q, R]$$

Figures (1),(2) show trajectories of the noisy model(14) and estimated value $\hat{x} = [\hat{x}'_1, \hat{x}'_2, \hat{x}'_3]^T$ with EKF in time domain and phase domain respectively. Value of the error is $error(x', \hat{x}') = 0.0017$ in simulation.

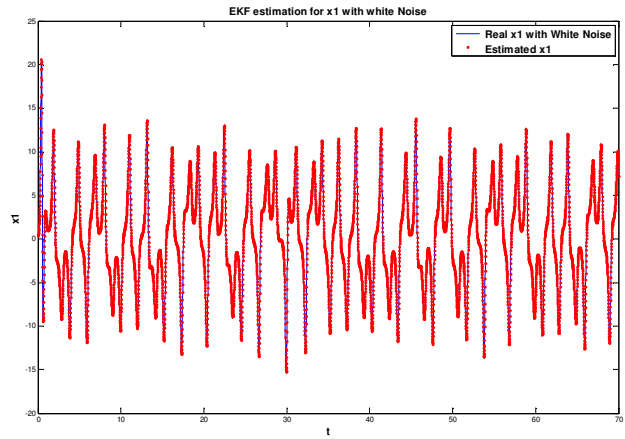

Fig.7 $x'_1$ for system (13) and $\hat{x}'_1$ estimated value with EKF

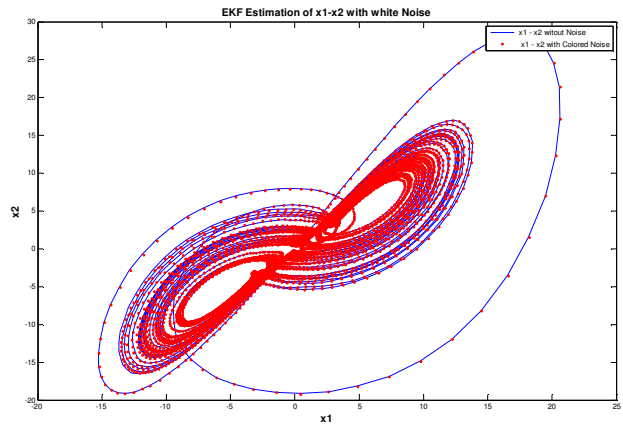

Fig.8 $x'_1 - x'_2$ for system (13) and $\hat{x}'_1 - \hat{x}'_2$ estimated values with EKF

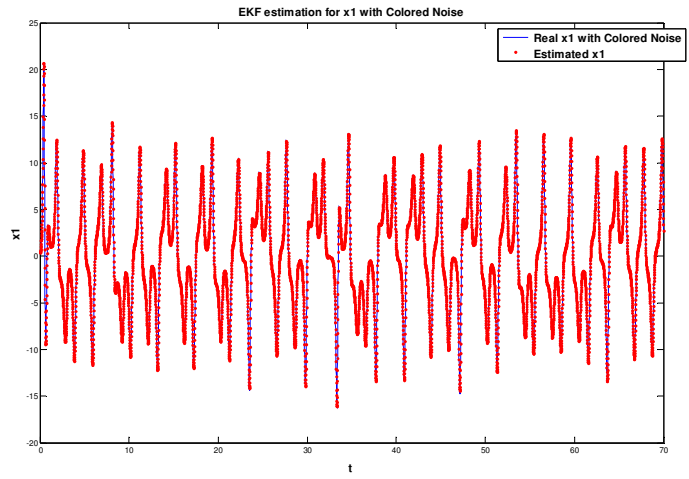

Fig.9 $x_1''$ for system (14) and $\hat{x}_1''$ estimated value with EKF

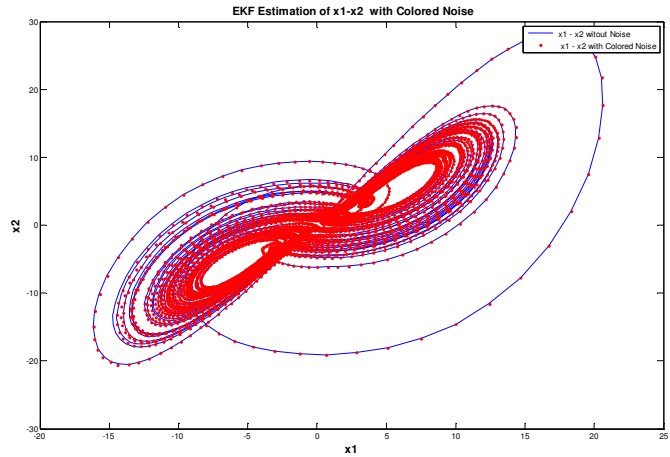

Fig.10 $x_1'' - x_2''$ for system (14) and $\hat{x}_1'' - \hat{x}_2''$ estimated values with EKF

## 5.3 EKF for state estimation of system with colored noise and unknown parameters

Realization of the Lorenz system with white noise and uncertain parameters results in

$$x'''_{1k+1} = x'''_{1k} + \Delta t \bar{\sigma}(x'''_{2k} - x'''_{1k}) + w_k$$
$$x'''_{2k+1} = x'''_{2k} + \Delta t(-x'''_{1k}x'''_{3k} + \bar{r}x'''_{1k} - x'''_{2k}) + w_k$$
$$x'''_{3k+1} = x'''_{3k} + \Delta t(x'''_{1k}x'''_{2k} - \bar{b}x'''_{3k}) + w_k$$
$$w_{k+1} = cw_k + e^1 \tag{16}$$
$$v_{k+1} = dv_k + e^2$$
$$z = x'''_{1k} + x'''_{3k} + v_k$$
$$e^1 \sim N(0,Q^1), e^2 \sim N(0,Q^2),$$
$$\bar{\sigma} = \sigma \pm \Delta\sigma, \bar{b} = b \pm \Delta b, \bar{r} = r \pm \Delta r, \bar{c} = c \pm \Delta c, \bar{d} = d \pm \Delta d$$

For realization in the form (7) parameters are
$$x = [x'''_1, x'''_2, x'''_3, v, w, c, d, \sigma, b, r]^T, \theta = []^T,$$
$$Q' = [Q^1, Q^2, \Delta c, \Delta d, \Delta\sigma, \Delta b, \Delta r]$$

Figures (11), (12) show trajectories of the noisy model(16) and estimated value with EKF in time domain and phase domain respectively. According to figures EKF estimated the real value of $x'''$ correctly. Value of the error is $error(x'''_1, \hat{x}'''_1) = 0.0941$ in simulation. Figure (13) shows estimated value $[\hat{\sigma}, \hat{b}, \hat{r}]$ with EKF. According to figures EKF estimated the real value of the system parameters correctly. Value of the error is $error(\theta, \hat{\theta}) = 0.1359$ in simulation.

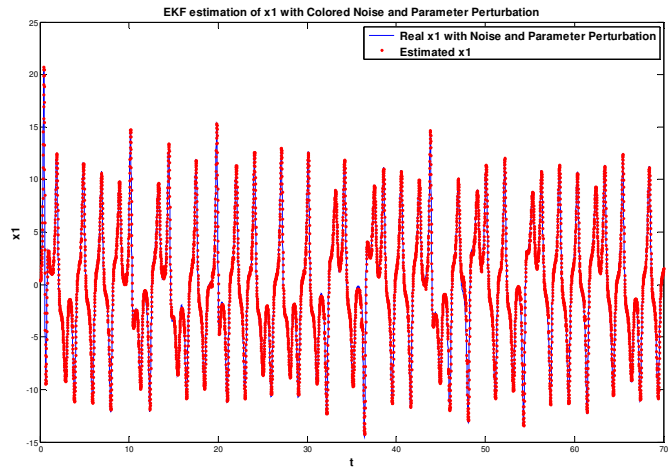

Fig.11 $x_1'''$ for system (16) and $\hat{x}_1'''$ estimated value with EKF

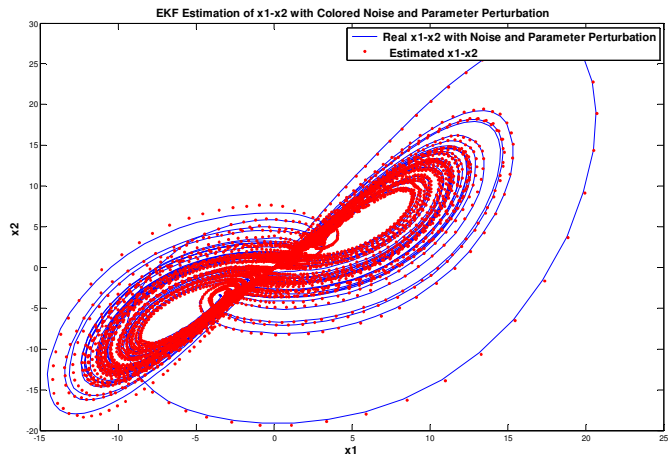

Fig.12 $x_1''' - x_2'''$ for system (16) and $\hat{x}_1''' - \hat{x}_2'''$ estimated values with EKF

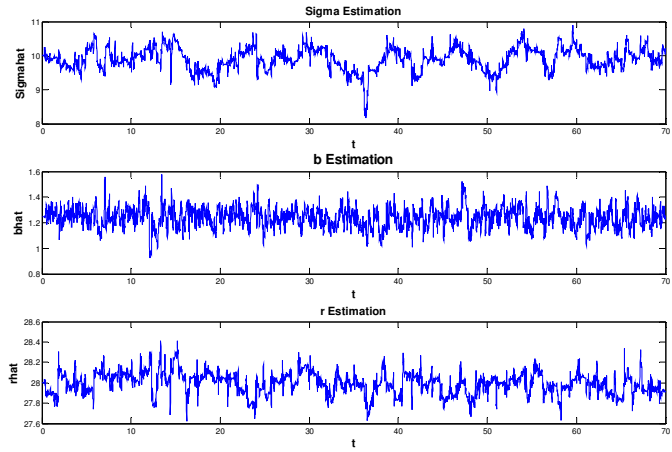

Fig.13 $\hat{\sigma}, \hat{b}, \hat{r}$ estimation for system (16) with EKF

## 6. Conclusion

The simulation results satisfying the efficiency of the EKF , a the first order filter which is result of Taylor expansion and neglecting higher order terms of the derivatives of the system equations, for estimation of the states and the parameters of the Lorenz system which is highly nonlinear chaotic system with noise like behavior.

## Refrences